\documentclass[twocolumn,showpacs,preprintnumbers,amsmath,amssymb,aps,pra,groupedaddress,showpacs]{revtex4-2}
\usepackage{multirow,diagbox}  
\usepackage{siunitx}

\usepackage[colorlinks,citecolor=blue,linkcolor=red,hyperindex,CJKbookmarks]{hyperref}
\usepackage{amsmath}
\usepackage{graphicx}
\usepackage{dcolumn}
\usepackage{subfigure}
\usepackage{mathrsfs}
\usepackage{amssymb}
\usepackage{amsfonts}
\usepackage{makecell}
\usepackage[dvipsnames]{xcolor}

\begin{document}
\draft
\title{Quantum Geometric Tensor and Critical Metrology in the Anisotropic Dicke Model}
\author{Xin Zhu, Jia-Hao L\"{u}, Wen Ning, Li-Tuo Shen}
\author{Fan Wu}\thanks{E-mail: t21060@fzu.edu.cn}
\author{Zhen-Biao Yang}\thanks{E-mail: zbyang@fzu.edu.cn}
\affiliation{Fujian Key Laboratory of Quantum Information and Quantum Optics,\\
	College of Physics and Information Engineering, Fuzhou University, Fuzhou,\\
	Fujian 350108, China}
\date{\today }
\date{\today }
\begin{abstract}

We investigate the quantum phase transition in the anisotropic Dicke model through an examination of the quantum geometric tensor of the ground state.  In this analysis, two distinct classical limits exhibit their unique anisotropic characteristics.  The classical spin limit demonstrates a preference for the rotating-wave coupling, whereas the classical oscillator limit exhibits symmetry in the coupling strength of the bias.  The anisotropic features of the classical spin limit persist at finite scales.  Furthermore, we observe that the interplay among the anisotropic ratio, spin length, and frequency ratio can collectively enhance the critical behaviors. This critical enhancement without trade-off between these factors provides a flexible method for  quantum precision measurement.
\end{abstract}

\date{\today}

\keywords{quantum sensing, quantum Fisher information, anisotropic quantum Rabi model}
\pacs{}

\maketitle

\section{INTRODUCTION}
The Dicke model~\cite{dicke} describes the interaction between $\mathcal{N}$ identical two-level systems and a single-mode bosonic field via a dipole coupling, which is of key importance as a model describing the collective effects in quantum optics~\cite{dicke1,dicke2}. A huge body of literature has been built around the Dicke model. It is useful for understanding nonequilibrium dynamics~\cite{nqd1,nqd2,nqd3,nqd4,nqd5,nqd6,nqd7} and studying the ultrastrong coupling regime in quantum systems~\cite{usc1,usc2,usc3,usc4}. It is also a paradigmatic model to benchmark tools detecting quantum chaos~\cite{chaos1,chaos2,chaos3,chaos4}. Recently it was employed to analyze the relationship between classical chaos and the evolution of out-of-time-ordered correlators~\cite{otoc1,otoc2,otoc3} and the presence of quantum scars~\cite{qs1,qs2}. In the classical spin (CS) limit with the spin length $\mathcal{N}\to\infty$ it exhibits a normal-to-superradiant quantum phase transition (QPT) at the critical point, which was first studied by Hepp and Lieb~\cite{qpt} at weak coupling. For the finite $\mathcal{N}$, the Dicke model becomes nonintegrable due to the unclosed Hilbert space in a general coupling range~\cite{range}. Consequently, several approximate methods have been devised to capture the effect of finite-size corrections~\cite{fs1,fs2,fs3,fs4}, proving pivotal in comprehending the universal properties near the critical point of QPT~\cite{qptp1,qptp2,qptp3,qptp4}.

A generalized version of the Dicke model, namely, the anisotropic Dicke model (ADM) ($\hbar = 1$), 
\begin{eqnarray}\label{adm}
	\mathcal{H} &=&\omega a^{\dagger}a + \Omega J_z + \frac{\lambda_1}{\sqrt{\mathcal{N}}}(e^{i\theta}a^{\dagger}J_- + e^{-i\theta}aJ_+) \nonumber\\
	&& + \frac{\lambda_2}{\sqrt{\mathcal{N}}}(e^{i\theta}a^{\dagger}J_+ + e^{-i\theta}aJ_-),
\end{eqnarray}
where $\omega$ is the frequency of single-mode bosonic field described by the bosonic operators $a^{(\dagger)}$, $\Omega$ is the transition  frequency and $J_{\pm,z} = \sum_{i=1}^{2j}\frac{1}{2}\sigma_{\pm,z}^{i}$ are the collective spin operators with pseudospin $j = \mathcal{N}/2$. $\lambda_{1}$ and $\lambda_2$ are the amplitudes of the coupling strengths corresponding to the rotating-wave and counterrotating-wave  coupling terms, where an extra phase $\theta$ is introduced. When $\lambda_1=\lambda_2=\lambda$, the ADM reduces to the conventional Dicke model. While the properties of the Dicke model have been studied extensively, they are also prominently present in the ADM~\cite{adm1,adm2,adm3,adm4,adm5,adm6}, where the asymmetric coupling strength endows it with some new properties at the same time. QPTs are often investigated by a variety of characterizations for quantumness, wherein entanglement was the earliest and most famous one~\cite{entanglement1,entanglement2,entanglement3,entanglement4,entanglement5}. Quantum discord, another quantity that characterizes the quantum correlation in certain situations can be used to detect QPTs~\cite{qd1,qd2,qd3}. Quantum geometric tensor (QGT), as the  preponderant element for studying the geometry of the quantum parameter space, was also useful in characterizing QPTs~\cite{qgtb,qgtb1,qgtb2}. In addition, QPT also represents a powerful resource~\cite{res1,res2,res3,res4,res5} for quantum metrology, which provides techniques to enhance the precision of measurements of physical quantities.

In this article, we address a QPT in the classical oscillator (CO) limit with the frequency ratio $\eta = \Omega/\omega\to\infty$. In contrast to the CS limit, a QPT occurs here already at finite $\mathcal{N}$. We investigate the normal-to-superradiant QPT behavior of the ADM and its dependence on the asymmetric coupling strengths in both limits. First, we observe the QPT in the ADM utilizing the quantum metric tensor (QMT) and the Berry curvature. The CS and CO limits give rise to QPTs with identical critical behaviors based on the mean-field theory~\cite{limit}. Furthermore, we assess the metrological potential of a finite-component QPT by measuring the Quantum Fisher Information (QFI)~\cite{QFI}, which is directly related to metrological sensitivity by the quantum Cram\'{e}r-Rao bound $\delta_A \ge 1/\sqrt{\mathcal{M}\mathcal{I}_A}$~\cite{QCRB}, where $\mathcal{M}$ is the number of measurements and $\mathcal{I}_A =  4[\langle\partial_A\psi\vert\partial_A\psi\rangle + (\langle\partial_A\psi\vert\psi\rangle)^2]$ is the QFI, where $\vert\psi\rangle$ denotes the arbitrary state of the system and $\partial_A = \frac{\partial}{\partial A}$. By comparing the QFI of the ground state under different anisotropic ratios, we find that both limits can be distinguished through the dependence of the anisotropic ratio. In the CS limit, the rotating-wave coupling preponderates more to QPT than the counterrotating-wave coupling does. But the contributions of the rotating-wave and counterrotating-wave interaction terms are symmetric in the CO limit. At finite scales, the QFI exhibits the same anisotropic ratio dependence as in the CS limit. Through contrasting anisotropic ratio, spin length, and frequency ratio that affect QFI, we are pleasantly surprised to find that they can be varied synchronously to make the QFI closer to the thermodynamic limit than when varied individually, which means that they can complement each other for better measurement precision under limited conditions.

\section{QUANTUM GEOMETRIC TENSOR}
We briefly review the concept of the QGT which characterizes the
geometry of a quantum system. Consider two quantum states $\vert\psi(p)\rangle$ and $\vert\psi(p + \delta p)\rangle$ that differ infinitesimally in a set of parameters $p = \left\{p_i\right\}(i = 1,2,\dots,M)$, the components of the QGT are defined as
\begin{eqnarray}
	\mathcal{Q}_{\mu\upsilon} = \langle\partial_\mu\psi\vert\partial_\upsilon\psi\rangle - \langle\partial_\mu\psi\vert\psi\rangle\langle\psi\vert\partial_\upsilon\psi\rangle,
\end{eqnarray}
which can take complex values. The real part of QGT is the QMT, denoted as $\mathcal{G}_{\mu\upsilon} = Re[\mathcal{Q}_{\mu\upsilon}]$, which is symmetric and defines a distance $ds^2 = 1 - \vert\langle\vert\psi(p)\vert\psi(p + \delta p)\rangle\vert^2 = \sum_{\mu\upsilon}^{} \mathcal{G}_{\mu\upsilon}dp_\mu dp_\upsilon$ between two nearby states $\vert\psi(p)\rangle$ and $\vert\psi(p + \delta p)\rangle$~\cite{qmtb}. The imaginary part is related to the Berry curvature $\mathcal{F}_{\mu\upsilon} = 2Im[\mathcal{Q}_{\mu\upsilon}] = -\mathcal{F}_{\upsilon\mu}$~\cite{berry}, which is antisymmetric and can be used to synthesize the first Chern number~\cite{chen1,chen2,chen3,chen4} after being integrated over a surface subtended by a closed path in the parameter space.

In particular, if we initialize the system in a nondegenerate eigenstate $\vert\psi(p)\rangle = \vert n(x)\rangle$, with the corresponding nondegenerate eigenvalue $E_n(p)$, as long as the parameter is slowly varying functions of time, the adiabatic theorem guarantees that the system will remain in this eigenstate during the evolution. The QMT can be expanded in a simple perturbative argument as~\cite{qgt}
\begin{eqnarray}\label{qmt}
	\mathcal{Q}_{\mu\upsilon}^{(n)} = \sum_{m\neq n}\frac{\langle n\vert\partial_\mu H\vert m \rangle\langle m\vert\partial_\upsilon H\vert n \rangle}{(E_m-E_n)^2}.
\end{eqnarray}
As mentioned above, this metric tensor characterizes the distance between two states, meaning that the larger its value, the more prominent statistical distinguishability of the two states. Eq.~(\ref{qmt}) shows that at the stationary points of the QPT, which are characterized by the ground-state level crossing, the components of the QGT are singular.

\section{THE QPT OF TWO DIFFERENT LIMITS}\label{two}
The ADM is invariant under the unitary
transformation $\Pi = \exp\left\{i\pi[a^{\dagger}a + J_z + j]\right\}$, such that $[\mathcal{H},\Pi] = 0$. The eigenvalues of $\Pi$ are $\pm1$, depending on whether the number of
qubit is even or odd, which allows the separation of the Hamiltonian matrix into two subspaces of definite parity. Unless stated, we work exclusively in the positive parity subspace.

The ADM undergoes a second-order QPT at $g = \frac{\lambda_1+\lambda_2}{\sqrt{\omega\Omega}}=1$ in both the CS and CO limits, separating  the system into two phases: the normal phase for $g<1$, and the superradiant phase for $g> 1$, respectively. 

To calculate the effective Hamiltonian of the CS limit, we use the Holstein-Primakoff representation of the angular momentum operators, which represent the operators in terms of a single-mode bosonic in the following way~\cite{HPT}:
\begin{eqnarray}\label{hp}
	J_+&=&b^{\dagger}\sqrt{2j-b^{\dagger}b},\nonumber\\
	J_- &=& \sqrt{2j-b^{\dagger}b}b, \nonumber\\
	J_z&=& b^{\dagger}b-j,
\end{eqnarray}
where $b$ and $b^{\dagger}$ are bosonic operators obeying $[b,b^{\dagger}] = 1$. This thus converts $\mathcal{H}$ into a two-mode bosonic problem. 

In this representation, the parity operator $\Pi$ becomes $\Pi = \exp\left\{i\pi[a^{\dagger}a + b^{\dagger}b]\right\}$. To consider the CS limit, we expand the square roots in Eq~.(\ref{hp}) and neglect terms with powers of $j$ in the denominator, which leads to 
\begin{eqnarray}
	J_+ &\simeq& \sqrt{2j}b^{\dagger},\nonumber\\
	J_- &\simeq& \sqrt{2j}b,\nonumber\\
	J_z &=& b^{\dagger}b-j.
\end{eqnarray}
In the normal phase $g<1$, this yields the effective Hamiltonian as 
\begin{eqnarray}\label{sn}
	\mathcal{H}^{cs}_{np} &\simeq& \omega a^{\dagger}a + \Omega(b^{\dagger}b - j) + \lambda_1(e^{i\theta}a^{\dagger}b + e^{-i\theta}ab^{\dagger}) \nonumber\\
	&& + \lambda_2(e^{i\theta}a^{\dagger}b^{\dagger} + e^{-i\theta}ab),
\end{eqnarray}
which is bilinear in the bosonic operators.

In the superradiant phase $g>1$, we need to displace the Hamiltonian, which can be achieved by rotating the angular momentum operators as
\begin{eqnarray}
	\left(\begin{array}{ccc}
		J_x \\
		J_y \\
		J_z
	\end{array}\right) = \left(\begin{array}{ccc}
	\cos\delta & 0 & \sin\delta \\
	0 & 1 & 0 \\
	-\sin\delta & 0 & \cos\delta
\end{array}\right)\left(\begin{array}{ccc}
		J_x^{'} \\
J_y^{'} \\
J_z^{'}
\end{array}\right),
\end{eqnarray}
and displacing the bosonic operators as
\begin{eqnarray}
	a = a^{'} + \alpha^{*}, a^{\dagger} = a^{'\dagger} + \alpha,
\end{eqnarray}
where $\delta$ and $\alpha$ are to be determined. We apply the truncated Holstein-Primakoff
transformation to the rotated operators $J^{'}_{x,y,z}$, i.e.,
\begin{eqnarray}
	J_x^{'} &\simeq& \sqrt{\frac{j}{2}}(b^{'\dagger}+b^{'}), \nonumber\\
	J_y^{'} &\simeq& -i\sqrt{\frac{j}{2}}(b^{'\dagger}-b^{'}), \nonumber\\
	J_z^{'} &=& b^{'\dagger}b^{'}-j,
\end{eqnarray}
to find the values of $\delta$ and $\alpha$. The Hamiltonian becomes
	\begin{eqnarray}\label{hsp1}
		\mathcal{H}_{sp} &\simeq& \omega\vert\alpha\vert^2 + \omega a^{'\dagger}a^{'} + \Omega\cos\delta(b^{'\dagger}b^{'}-j) \nonumber\\
		&&+\frac{\lambda_1+\lambda_2}{2}\cos\delta(e^{i\theta}a^{'\dagger} + e^{-i\theta}a^{'})(b^{'\dagger}+b^{'}) \nonumber\\
		&&+ \sqrt{\frac{2}{j}}(\lambda_1+\lambda_2)\vert\alpha\vert\sin\delta(b^{'\dagger}b^{'}-j)  \nonumber\\
		&&+\frac{\lambda_1+\lambda_2}{\sqrt{2j}}\sin\delta(e^{i\theta}a^{'\dagger} + e^{-i\theta}a^{'})b^{'\dagger}b^{'} \nonumber\\
		&&+ \left[\omega\vert\alpha\vert - \sqrt{\frac{j}{2}}(\lambda_1+\lambda_2)\sin\delta\right](e^{i\theta}a^{'\dagger} + e^{-i\theta}a^{'}) \nonumber\\
		&&+ \left[(\lambda_1+\lambda_2)\vert\alpha\vert\cos\delta - \Omega\sqrt{\frac{j}{2}}\sin\delta\right](b^{'\dagger}+b^{'}) \nonumber\\
		&&-\frac{\lambda_1-\lambda_2}{2}(e^{i\theta}a^{'\dagger} - e^{-i\theta}a^{'})(b^{'\dagger} - b^{'}).
	\end{eqnarray}
Now we eliminate the terms in Eq.~(\ref{hsp1}) that are linear in $(a^{'\dagger} + a^{'})$ and $(b^{'\dagger}+b^{'})$ by choosing the values of $\alpha$ and $\delta$, so that
\begin{eqnarray}
	\omega\vert\alpha\vert - \sqrt{\frac{j}{2}}(\lambda_1+\lambda_2)\sin\delta = 0, \nonumber\\
	(\lambda_1+\lambda_2)\vert\alpha\vert\cos\delta - \Omega\sqrt{\frac{j}{2}}\sin\delta = 0.
\end{eqnarray}
The nontrivial solution gives
\begin{eqnarray}\label{alpha}
		&\alpha = \frac{e^{i\theta}\sqrt{2j[(\lambda_1+\lambda_2)^4 - \omega^2\Omega^2]}}{2(\lambda_1+\lambda_2)\omega}&, \nonumber\\ 
		&\cos\delta = \frac{\omega\Omega}{(\lambda_1+\lambda_2)^2}.&
\end{eqnarray}
With these determinations, we obtain the Hamiltonian as
	\begin{eqnarray}\label{hsp2}
		\mathcal{H}^{cs}_{sp} &\simeq& \frac{\omega\Omega}{2(\lambda_1+\lambda_2)}(e^{i\theta}a^{'\dagger} + e^{-i\theta}a^{'})(b^{'\dagger}+b^{'}) \nonumber\\
		&&-\frac{\lambda_1-\lambda_2}{2}(e^{i\theta}a^{'\dagger} - e^{-i\theta}a^{'})(b^{'\dagger} - b^{'}) \nonumber\\
		&& +\omega a^{'\dagger}a^{'} + \frac{(\lambda_1+\lambda_2)^2}{\omega}b^{'\dagger}b^{'} \nonumber\\
		&&-j\left[\frac{(\lambda_1+\lambda_2)^2}{2\omega}+\frac{\omega\Omega^2}{2(\lambda_1+\lambda_2)^2}\right]  .
	\end{eqnarray}	
This effective Hamiltonian $\mathcal{H}^{cs}_{sp}$ in the superradiant phase is also bilinear with the bosonic operators. Although the global symmetry $\Pi$ becomes broken at the phase transition, two new local symmetries appear, corresponding to the operator $\Pi = \exp\left\{i\pi[a^{'\dagger}a^{'} + b^{'\dagger}b^{'}]\right\} $.

For the CO limit, we apply Schrieffer–Wolff transformation to obtain an effective low-energy Hamiltonian. The transformed Hamiltonian is written as
\begin{eqnarray}\label{sh}
	\mathcal{H}^{'} = \exp^{-S}\mathcal{H}\exp^{S} = \sum_{k = 0}^{\infty}\frac{[\mathcal{H},S]^{k}}{k!},
\end{eqnarray}
where $[\mathcal{H},S]^{k} = [[\mathcal{H},S]^{k-1},S]$ with $[\mathcal{H},S]^{0} = \mathcal{H}$. We divide Eq.~(\ref{sh}) into diagonal and off-diagonal parts, where we have denoted the special anti-Hermitian operator $S$ as
\begin{eqnarray}
	&S = S_1 + S_2,&
\end{eqnarray}
with
\begin{eqnarray}
		&S_1 = -\frac{1}{\sqrt{2j}\Omega}(AJ_+ -BJ_-) + \mathcal{O}\left(\omega/\Omega^2\right),&\nonumber\\
	&S_2 = \frac{4}{3(\sqrt{2j}\Omega)^3}(ABAJ_+ - BABJ_-)+ \mathcal{O}\left(\omega/\Omega^4\right),&
\end{eqnarray}
in which $A = \lambda_1e^{-i\theta}a + \lambda_2e^{i\theta}a^{\dagger}$ and $B = \lambda_1e^{i\theta}a^{\dagger} + \lambda_2e^{-i\theta}a$.

Projecting Eq.~(\ref{sh}) into ${\vert j,-j\rangle}$, i.e., $\mathcal{H}^{co}_{np} \equiv\langle j,-j\vert\mathcal{H}^{'}\vert j,-j\rangle$, and  keeping the terms up to the first order in $\omega/\Omega$, it results in an effective low-energy Hamiltonian
\begin{eqnarray}\label{on}
	\mathcal{H}^{co}_{np} &\simeq& - \frac{1}{\Omega}(\lambda_1 e^{i\theta}a^{\dagger} + \lambda_2 e^{-i\theta}a)(\lambda_1e^{-i\theta}a + \lambda_2e^{i\theta}a^{\dagger}) \nonumber\\
	&& + \omega a^{\dagger}a - j\Omega.
\end{eqnarray}

For $g>1$, we need to displace the Hamiltonian $\mathcal{H}$ as previously done within the CS limit, employing identical formulas for $\alpha$ and $\delta$ as Eq.~(\ref{alpha}). Therefore $\mathcal{H}$ approximates to be
\begin{eqnarray}
	\tilde{\mathcal{H}} &\simeq& \omega a^{' \dagger}a^{'} + \tilde{\Omega} J_z^{'} +\frac{\lambda_1^{'}}{\sqrt{2j}}(e^{i\theta}a^{' \dagger} J_-^{'} + e^{-i\theta}a^{'}J_+^{'}) \nonumber\\
	&&+ \frac{\lambda_2^{'}}{\sqrt{2j}}(e^{i\theta}a^{' \dagger} J_+^{'} + e^{-i\theta}a^{'}J_-^{'}) \nonumber\\
	&&+ \frac{j\Omega}{2}\left(g^2 - g^{-2}\right),
\end{eqnarray}
with the rescaled frequency and qubit-oscillator coupling strengths being 
\begin{eqnarray}
	&\tilde{\Omega} = \Omega g^2,&\nonumber\\
	&\lambda_1^{'} = \frac{1}{2\sqrt{2j}}\left[\sqrt{\omega\Omega}g^{-1} + (\lambda_1 -\lambda_2)\right],&\nonumber\\
	&\lambda_2^{'} = \frac{1}{2\sqrt{2j}}\left[\sqrt{\omega\Omega}g^{-1} - (\lambda_1 -\lambda_2)\right].&
\end{eqnarray}
With the application of the same method described above in the normal phase,  the effective Hamiltonian for the superradiant phase reaches as follows:
\begin{eqnarray}\label{os}
	\mathcal{H}^{co}_{sp} &\simeq&  - \frac{1}{\tilde{\Omega}}(\lambda_1^{'}e^{i\theta}a^{'\dagger} + \lambda_2^{'}e^{-i\theta}a^{'})(\lambda_2^{'}e^{i\theta}a^{'\dagger} + \lambda_1^{'}e^{-i\theta}a^{'})\nonumber\\
	&& + \omega a^{'\dagger}a^{'} - \frac{j\Omega}{2}\left(g^2 + g^{-2}\right).
\end{eqnarray}
The effective Hamiltonian $\mathcal{H}^{co}_{sp}$ has a similar structure to $\mathcal{H}^{co}_{np}$, which means that both phases exhibit analogous critical behaviors.

\begin{figure}[t]
	\centering
	\includegraphics[width=0.5\textwidth]{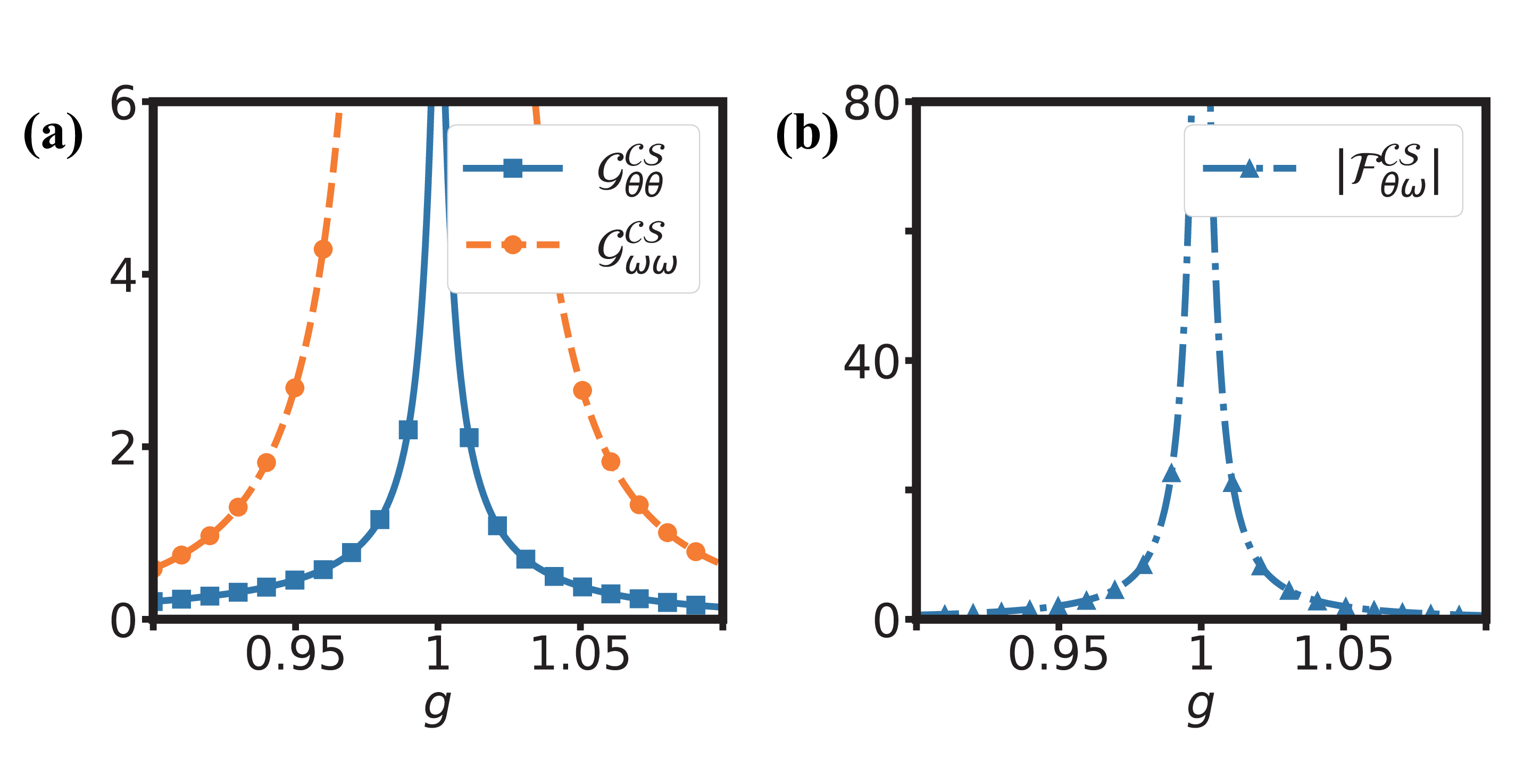}
	\includegraphics[width=0.5\textwidth]{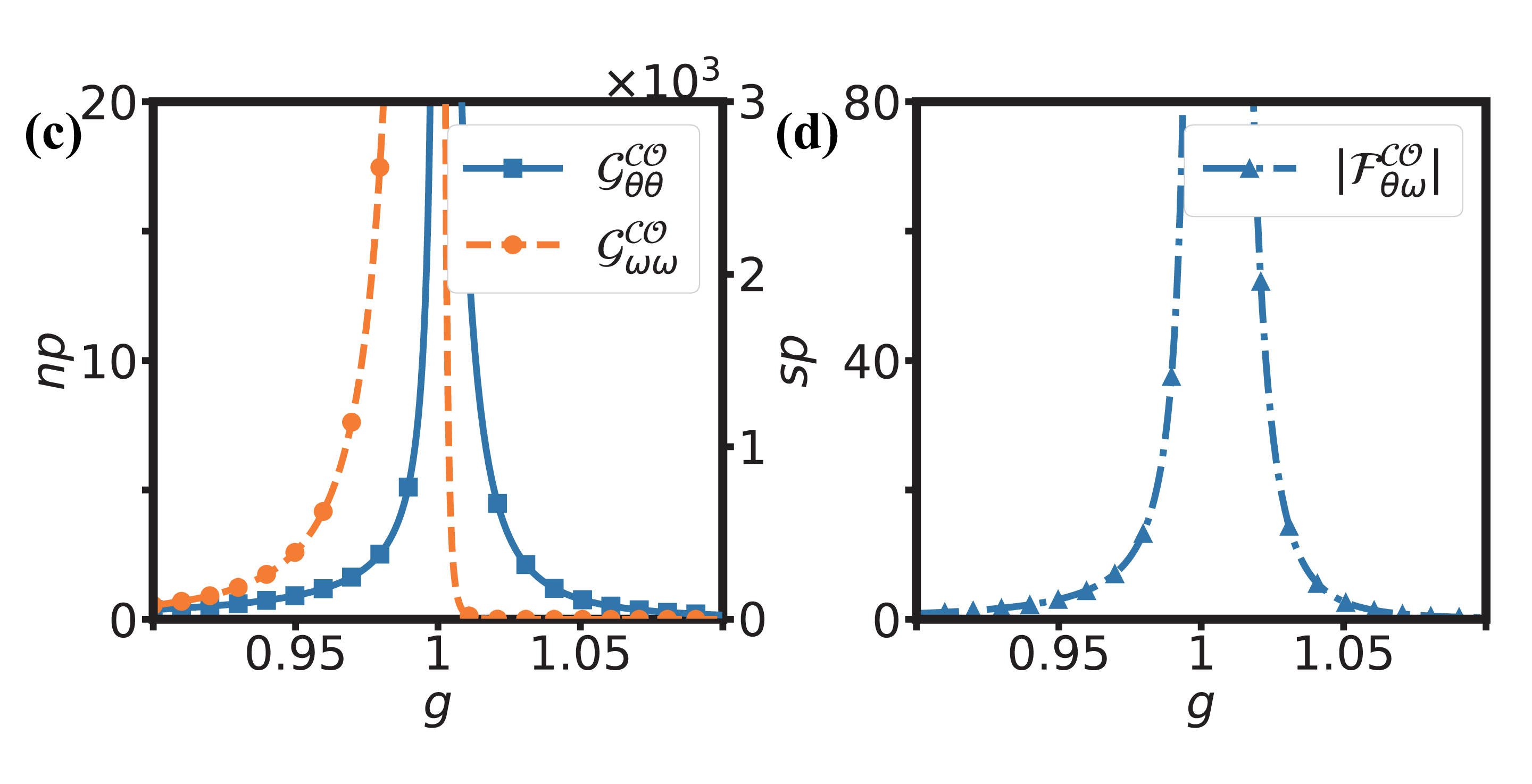}
	\caption{(a),(c) The QMT components as a function of $g$. (b),(d) The Berry curvature component as a function of $g$. In the CS limit with fixed $\eta = 1$ and in the 
		CO limit with fixed $j = 10$. The numerical results are obtained according to Eqs.~(\ref{sn}), (\ref{hsp2}), (\ref{on}), (\ref{os}).}
	\label{f1}
\end{figure}
\section{CRITICAL METROLOGY THROUGH QUANTUM GEOMETRIC TENSOR}

Based on the effective Hamiltonian in Sec.~\ref{two}, we expect that the components of both the QMT and Berry curvature show a divergent feature close to the critical point. In Fig.~\ref{f1}, we show the corresponding components from a numerical
calculation of the ground state of the ADM with $\mu = \theta$, $\nu = \omega$, $\gamma = \lambda_1/\lambda_2 = 2$, and the bosonic cutoff $n_{max}=100$. It can be seen from Fig.~\ref{f1} that, in both situations, the QMT and Berry curvature components exhibit singularity at the critical point. It is noted that the Berry curvature component has different divergence directions in the normal and superradiant phases due to the fact that the squeezing parameter has different formulation for $g$ (see Appendix~\ref{A} for a proof).

\begin{figure}[hbpt!]
	\centering
	\includegraphics[width=0.49\textwidth]{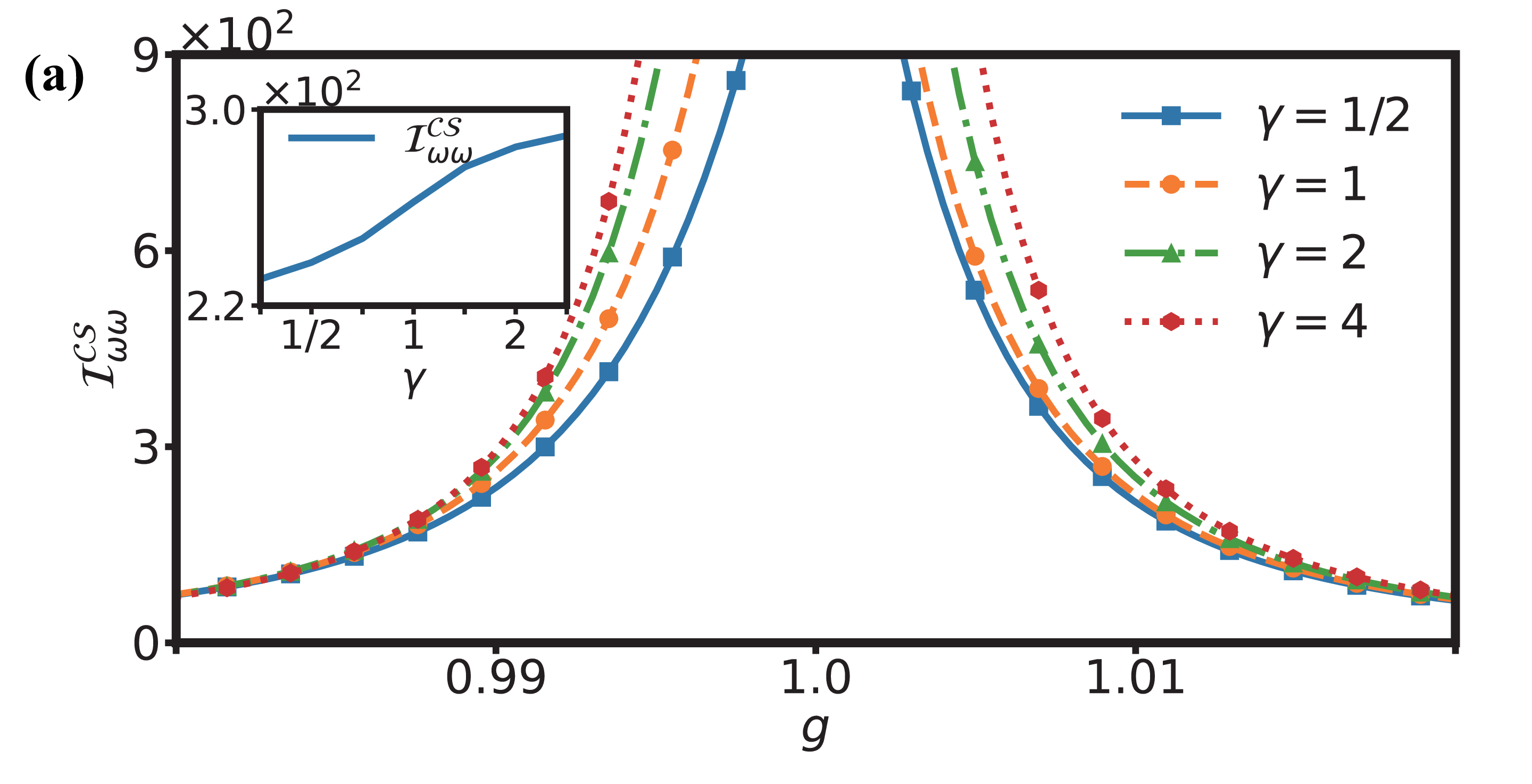}
\includegraphics[width=0.49\textwidth]{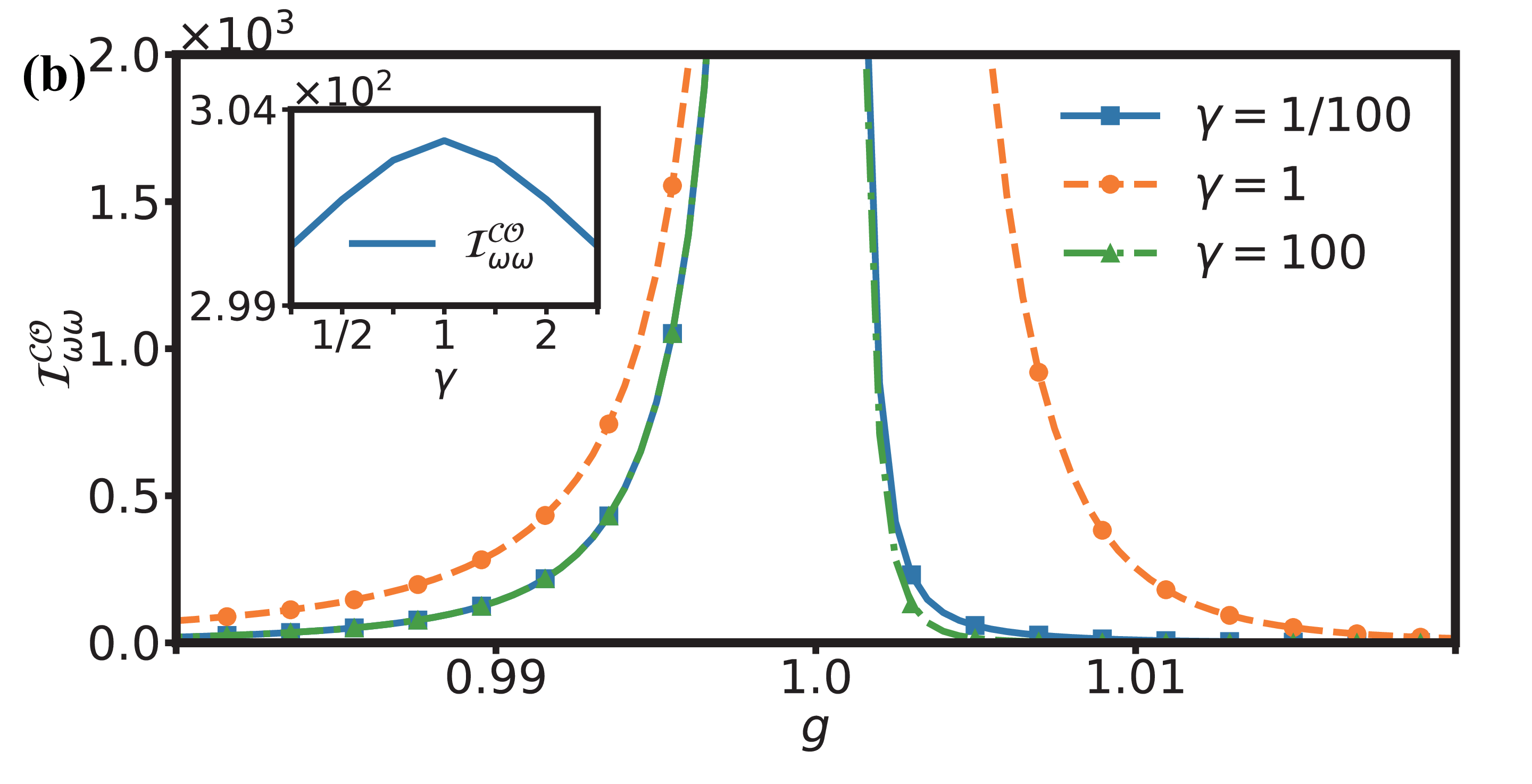}
	\caption{(a) The QFI $\mathcal{I}_{\omega\omega}$ as a function of $g$ with different $\gamma$ in the CS limit with fixed $\eta = 1$. The inset shows the variation of $\mathcal{I}_{\omega\omega}$ at $g = 0.99$.
	(b) The same data in the CO limit with fixed $j = 10$. The numerical results are obtained according to Eqs.~(\ref{sn}), (\ref{hsp2}), (\ref{on}), (\ref{os}).}
	\label{f2}
\end{figure}

Both the CS and CO limits present a critical behaviors of the QPT in Fig.~\ref{f1}. The nature of the  two limits is different, however, they can be distinguished through the dependence of $\gamma$. In Fig.~\ref{f2}, we compare the QFI $\mathcal{I}_{\omega\omega}$ (i.e., $4\mathcal{Q}_{\omega\omega}$) with different $\gamma$. From the inset, we can intuitively observe the dependence of the two limits on $\gamma$. In the CS limit, the Holstein-Primakoff transformation maintains the distinction between rotating-wave and counterrotating-wave interactions. Within this framework, rotating-wave interactions play a more favorable role in QPT compared to counterrotating-wave interactions. Conversely, in the CO limit, the projected Hamiltonian is free of coupling terms among the spin states. Consequently, this erases the distinction between rotating-wave and counterrotating-wave interactions, rendering them symmetric in their roles.

\begin{figure}[hbpt!]
	\centering
	\includegraphics[width=0.49\textwidth]{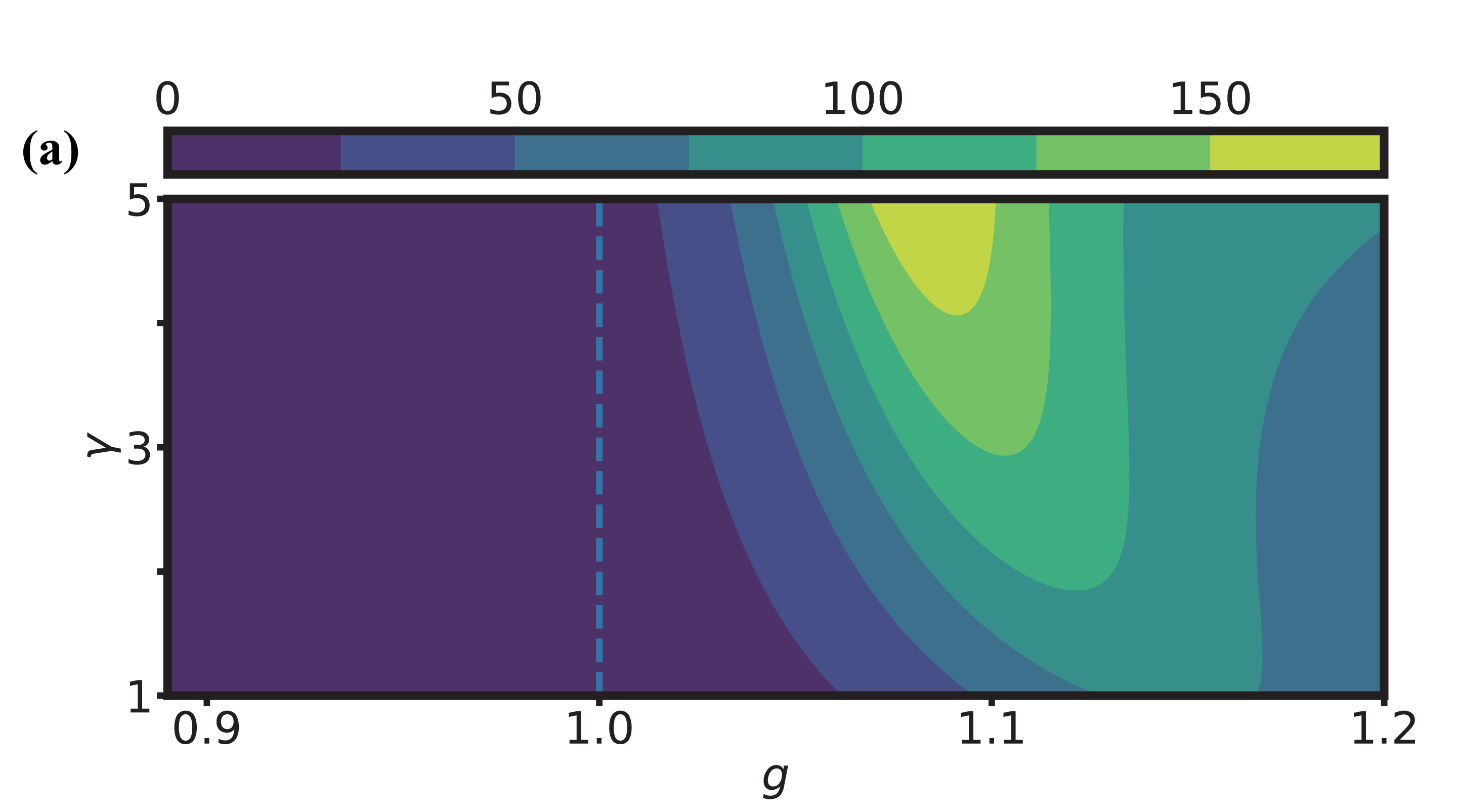}
\includegraphics[width=0.49\textwidth]{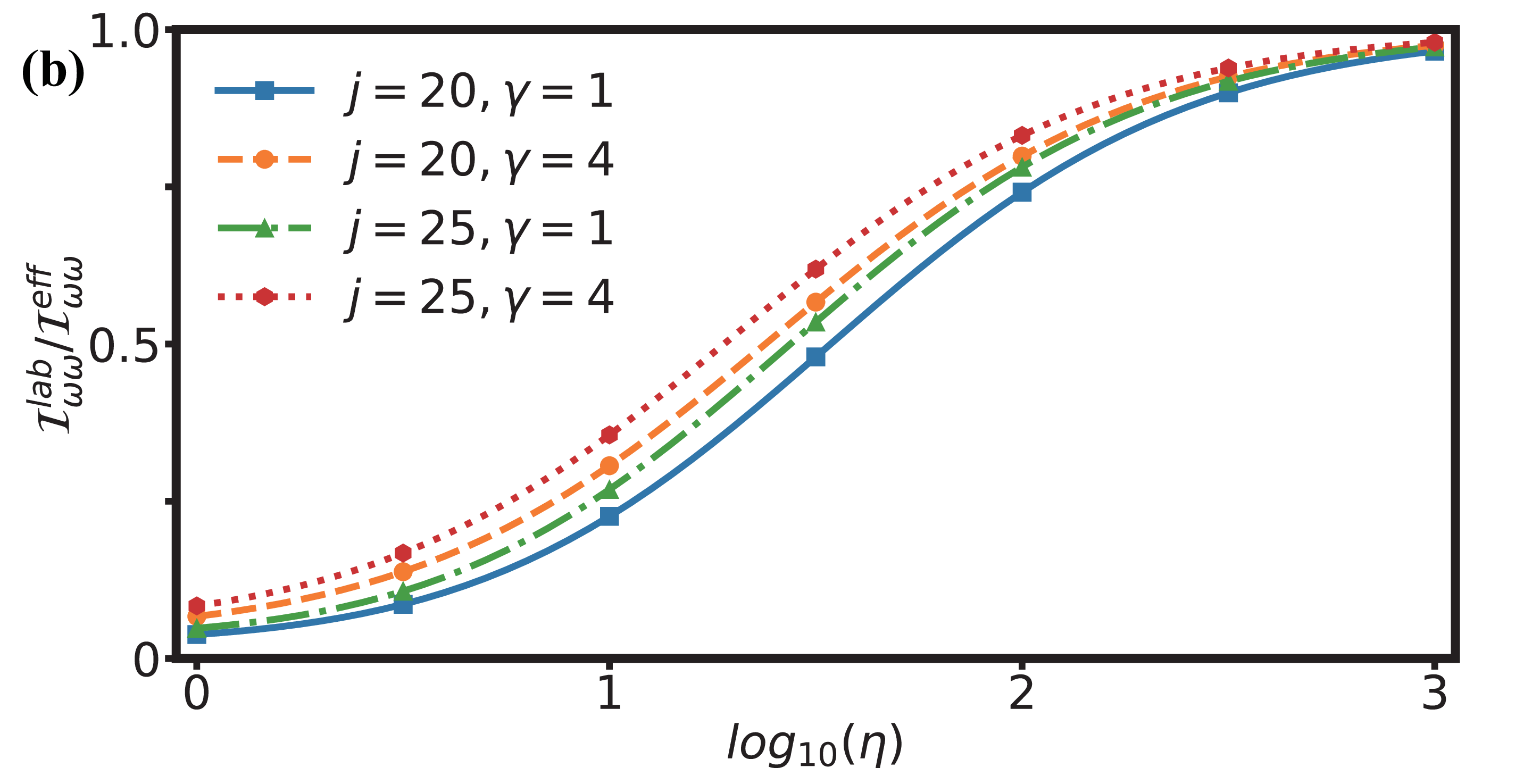}
	\caption{(a) The QFI of $\omega$ as a function of $g$ and $\gamma$ with fixed $j = 10$, $\eta = 1$. (b) The ratio of the QFI $\mathcal{I}_{\omega\omega}^{lab}/\mathcal{I}_{\omega\omega}^{eff}$ as a function of $\eta$ by selected numbers of $j$ and $\gamma$ at $g = 0.99$.}
	\label{f3}
\end{figure}

For finite $j$ and $\eta$, the variation of $\mathcal{I}_{\omega\omega}$ with different $\gamma$ is shown in Fig.~\ref{f3}(a), the dependence of $\mathcal{I}_{\omega\omega}$ on $\gamma$ is consistent with the case in the CS limit.
Fig.~\ref{f3}(b) shows the variation of the ratio of QFI $\mathcal{I}_{\omega\omega}^{lab}/\mathcal{I}_{\omega\omega}^{eff}$ with $\eta = \Omega/\omega$ in logarithms. The result indicates that the simultaneous adjustment of $j$, $\gamma$, and $\eta$ is more likely to push the QFI to approach the thermodynamic limit as compared to the case with individual adjustments, which means that we can optimize them simultaneously to improve QFI without having a trade-off between them.

\section{Conclusion}
As pointed out here that the QPT can be realized in two different classical limits for the ADM. Differences emerge in terms of their dependence on the anisotropic ratio. In the CS limit, the bias in the rotating-wave coupling facilitates the improvement of the QPT behavior of the system. In contrast, within the CO limit the influences of the rotating-wave and counterrotating-wave interaction terms are symmetric. Furthermore, the critical behavior at finite scales inherits the anisotropic dependency observed at the CS limit. Importantly, this bias coupling's effect harmonizes with the enhancement of spin length or frequency ratio, allowing us to enhance measurement precision at finite scales through multi-directional optimization. This approach offers valuable guidance for the practical implementation of quantum precision measurements at the current stage of scientific development.

\section{ACKNOWLEDGE}
This work was supported by the National Natural Science Foundation of
China (Grant Nos. 12274080 and 11875108), the National
Youth Science Foundation of China (Grant No. 12204105), the Educational
Research Project for Young and Middle-aged Teachers of Fujian Province
(Grant No. JAT210041), and the Natural Science Foundation of Fujian
Province (Grant Nos. 2021J01574, and 2022J05116).

\appendix\section{}\label{A}
For the Rabi model, the ground state in the normal phase is 
\begin{eqnarray}
	\vert\phi_{np}(g)\rangle = \mathcal{S}[r_{np}(g)]\vert0\rangle\left\vert\downarrow\right\rangle,
\end{eqnarray}
for which
\begin{eqnarray}
	&\mathcal{S}[r_{np}(g)] = \exp\left(1/2)[r_{np}^*(g)a^{2} - r_{np}(g)a^{\dagger2}\right]&, \\ &r_{np}(g) = e^{2i\theta}(1/4)\ln(1-g^2).&
\end{eqnarray}

While in the superradiant phase, the ground state reads 
\begin{eqnarray}
	\vert\phi_{sp}(g)\rangle =\mathcal{D}[\pm\alpha] \mathcal{S}[r_{sp}(g)]\vert0\rangle\left\vert\downarrow^{\pm}\right\rangle,
\end{eqnarray}
where 
\begin{eqnarray}
	&\mathcal{D}[\alpha] = \exp(\alpha a^{\dagger} - \alpha^* a),& \\
	  &r_{sp}(g) = e^{2i\theta}(1/4)\ln(1-g^{-4})&.
\end{eqnarray}

The corresponding Berry curvatures are
\begin{eqnarray}
	\mathcal{F}_{\theta\omega}^{np} &=& 2Im[- \langle\partial_\theta\phi_{np}(g)\vert\phi_{np}(g)\rangle\langle\phi_{np}(g)\vert\partial_\omega\phi_{np}(g)\rangle \nonumber\\
	&&+ \langle\partial_\theta\phi_{np}(g)\vert\partial_\omega\phi_{np}(g)\rangle] \nonumber\\
	&=& \vert r_{np}(g)\vert g^2\cosh(2\vert r_{np}(g)\vert)/\left[2\omega(1-g^2)\right],
\end{eqnarray}
and 
\begin{eqnarray}
	\mathcal{F}_{\theta\omega}^{sp} &=& 2Im[- \langle\partial_\theta\phi_{sp}(g)\vert\phi_{sp}(g)\rangle\langle\phi_{sp}(g)\vert\partial_\omega\phi_{sp}(g)\rangle \nonumber\\
	&&+ \langle\partial_\theta\phi_{sp}(g)\vert\partial_\omega\phi_{sp}(g)\rangle] \nonumber\\
	&=& -\vert r_{sp}(g)\vert \cosh(2\vert r_{sp}(g)\vert)/\left[g^4\omega(1-g^{-4})\right] \nonumber\\
	&&+ 2\lambda^2/\omega^3,
\end{eqnarray}
respectively. The second term of $\mathcal{F}_{\theta\omega}^{sp}$ can be neglected near the critical point.


\begin{thebibliography}{99}
\bibitem{dicke}R. H. Dicke, Phys. Rev. 93, 99 (1954).
\bibitem{dicke1}A.V. Andreev, V.J. Emel’yanov, and Yu.A. Il’inskii, Cooperative Effects in Optics (IOP, Bristol, 1993).
\bibitem{dicke2}M. G. Benedict, A. M. Ermolaev, V. A. Malyshev, I. V. Sokolov, and E. D. Trifonov, in \textit{Super-radiance: Multiatomic Coherent Emission}, 1st ed., edited by M. G. Benedict (CRC Press, Boca Raton, FL, 2018).
\bibitem{nqd1}P. P\'{e}rez-Fern\'{a}ndez, P. Cejnar, J. M. Arias, J. Dukelsky, J. E.
Garc\'{i}a-Ramos, and A. Rela\~{n}o, Phys. Rev. A 83, 033802 (2011).
\bibitem{nqd2}A. Altland and F. Haake, Phys. Rev. Lett. 108, 073601 (2012).
\bibitem{nqd3}H. Shen, P. Zhang, R. Fan, and H. Zhai, Phys. Rev. B 96, 054503 (2017).
\bibitem{nqd4}S. Lerma-Hern\'{a}ndez, J. Ch\'{a}vez-Carlos, M. A. BastarracheaMagnani, L. F. Santos, and J. G. Hirsch, J. Phys. A 51, 475302 (2018)
\bibitem{nqd5}S. Lerma-Hern\'{a}ndez, D. Villase\~{n}or, M. A. BastarracheaMagnani, E. J. Torres-Herrera, L. F. Santos, and J. G. Hirsch, Phys. Rev. E 100, 012218 (2019).
\bibitem{nqd6}M. Kloc, P. Str\'{a}nsk\'{y}, and P. Cejnar, Phys. Rev. A 98, 013836 (2018).
\bibitem{nqd7}P. Kirton, M. M. Roses, J. Keeling, and E. G. Dalla Torre, Adv. Quantum Technol. 2, 1800043 (2019).
\bibitem{usc1}D. De Bernardis, T. Jaako, and P. Rabl, Phys. Rev. A 97, 043820 (2018).
\bibitem{usc2}A. Frisk Kockum, A. Miranowicz, S. De Liberato, S. Savasta, and F. Nori, Nat. Rev. Phys. 1, 19 (2019).
\bibitem{usc3}P. Forn-D\'{i}az, L. Lamata, E. Rico, J. Kono, and E. Solano, Rev. Mod. Phys. 91, 025005 (2019).
\bibitem{usc4}A. Le Boit\'{e}, Adv. Quantum Technol. 3, 1900140 (2020).
\bibitem{chaos1}C. M. L\'{o}bez and A. Rela\~{n}o, Phys. Rev. E 94, 012140 (2016).
\bibitem{chaos2} L. Song, D. Yan, J. Ma, and X. Wang, Phys. Rev. E 79, 046220 (2009).
\bibitem{chaos3}P. Giorda and P. Zanardi, Phys. Rev. E 81, 017203 (2010).
\bibitem{chaos4}D. A. Wisniacki and A. J. Roncaglia, Phys. Rev. E 87, 050902(R) (2013)
\bibitem{otoc1}J. Ch\'{a}vez-Carlos, B. L\'{o}pez-del-Carpio, M. A. BastarracheaMagnani, P. Str\'{a}nsk\'{y}, S. Lerma-Hern\'{a}ndez, L. F. Santos, and J. G. Hirsch, Phys. Rev. Lett. 122, 024101 (2019).
\bibitem{otoc2}R. J. Lewis-Swan, A. Safavi-Naini, J. J. Bollinger, and A. M. Rey, Nat. Commun. 10, 1581 (2019).
\bibitem{otoc3} S. Pilatowsky-Cameo, J. Ch\'{a}vez-Carlos, M. A. BastarracheaMagnani, P. Str\'{a}nsk\'{y}, S. Lerma-Hern\'{a}ndez, L. F. Santos, and J. G. Hirsch, Phys. Rev. E 101, 010202(R) (2020).
\bibitem{qs1}D. Villase\~{n}or, S. Pilatowsky-Cameo, M. A. BastarracheaMagnani, S. Lerma-Hern\'{a}ndez, L. F. Santos, and J. G. Hirsch, New J. Phys. 22, 063036 (2020).
\bibitem{qs2}S. Pilatowsky-Cameo, D. Villase\~{n}or, M. A. BastarracheaMagnani, S. Lerma-Hern\'{a}ndez, L. F. Santos, and J. G. Hirsch, Nat. Commun. 12, 852 (2021).
\bibitem{qpt}K. Hepp and E. Lieb, Ann. Phys. (N. Y.) 76,360 (1973).
\bibitem{range}D. Braak, J. Phys. B: At., Mol. Opt. Phys. 46, 224007 (2013).
\bibitem{fs1}G. Liberti, F. Plastina, and F. Piperno, Phys. Rev. A 74, 022324 (2006).
\bibitem{fs2}J. Vidal and S. Dusuel, Europhys. Lett. 74, 817 (2006).
\bibitem{fs3}O. Tsyplyatyev and D. Loss, J. Phys. Conf. Ser. 193, 012134
(2009).
\bibitem{fs4}A. Rela\~{n}o, M. A. Bastarrachea-Magnani, and S. Lerma-Hern\'{a}ndez, Europhys. Lett. 116, 50005 (2016).
\bibitem{qptp1}N. Lambert, C. Emary, and T. Brandes, Phys. Rev. Lett. 92, 073602 (2004).
\bibitem{qptp2}W. Casteels, R. Fazio, and C. Ciuti, Phys. Rev. A 95, 012128 (2017).
\bibitem{qptp3}J. M. Fink, A. Dombi, A. Vukics, A. Wallraff, and P. Domokos, Phys. Rev. X 7, 011012 (2017).
\bibitem{qptp4}Q.-T. Xie, S. Cui, J.-P. Cao, L. Amico, and H. Fan, Phys. Rev. X 4, 021046 (2014).
\bibitem{adm1}W. Buijsman, V. Gritsev, and R. Sprik, Phys. Rev. Lett. 118, 080601 (2017).
\bibitem{adm2}M. Kloc, P. Str\'{a}nsk\'{y}, and P. Cejnar, Ann. Phys. (NY) 382, 85 (2017).
\bibitem{adm3}I. Aedo and L. Lamata, Phys. Rev. A 97, 042317 (2018)
\bibitem{adm4}J. Hu and S. Wan, Commun. Theor. Phys. 73, 125703 (2021).
\bibitem{adm5}K. Baumann, R. Mottl, F. Brennecke, and T. Esslinger, Phys. Rev. Lett. 107, 140402 (2011).
\bibitem{adm6}P. Das, D. S. Bhakuni, and A. Sharma, Phys. Rev. A 107, 043706 (2023)
\bibitem{entanglement1}L. Amico and D. Patane, Europhys. Lett. 77, 17001 (2007).
\bibitem{entanglement2} L. Amico, R. Fazio, A. Osterloh, and V. Vedral, Rev. Mod. Phys. 80, 517 (2008).
\bibitem{entanglement3}A. Osterloh, L. Amico, G. Falci, and R. Fazio, Nature (London) 416, 608 (2002).
\bibitem{entanglement4}T. J. Osborne and M. A. Nielsen, Phys. Rev. A 66, 032110 (2002).
\bibitem{entanglement5}G. Vidal, J. I. Latorre, E. Rico, and A. Kitaev, Phys. Rev. Lett. 90, 227902 (2003).
\bibitem{qd1}M. S. Sarandy, Phys. Rev. A 80, 022108 (2009).
\bibitem{qd2}T. Werlang, C. Trippe, G. A. P. Ribeiro, and G. Rigolin, Phys. Rev. Lett. 105, 095702 (2010).
\bibitem{qd3}G. Adesso and A. Datta, Phys. Rev. Lett. 105, 030501 (2010).
\bibitem{qgtb}D. Guti\'{e}rrez-Ruiz, J. Ch\'{a}vez-Carlos, D. Gonzalez, J. G. Hirsch and J. D. Vergara, Phys. Rev. B 105, 214106 (2022).
\bibitem{qgtb1}D. Guti\'{e}rrez-Ruiz, D. Gonzalez, J. Ch\'{a}vez-Carlos, J. G. Hirsch, and J. D. Vergara, Phys. Rev. B 103, 174104 (2021).
\bibitem{qgtb2}A. Dey, S. Mahapatra, P. Roy, and T. Sarkar, Phys. Rev. E 86, 031137 (2012).
\bibitem{res1}P. Zanardi, M. G. A. Paris, and L. Campos Venuti, Phys. Rev. A 78, 042105 (2008).
\bibitem{res2}Y. Chu, S. Zhang, B. Yu, and J. Cai, Phys. Rev. Lett. 126, 010502 (2021).
\bibitem{res3}L. Garbe, M. Bina, A. Keller, M. G. A. Paris, and S. Felicetti, Phys.
Rev. Lett. 124, 120504 (2020)
\bibitem{res4}J. H. L\"{u}, W. Ning, X. Zhu, F. Wu, L. T. Shen, Z. B. Yang and S. B. Zheng, Phys. Rev. A 106, 062616 (2022).
\bibitem{res5}X. Zhu, J. H. L\"{u}, W. Ning, F. Wu, L. T. Shen, Z. B. Yang and S. B. Zheng, Sci. China Phys. Mech. Astron. 66 250313 (2023).
\bibitem{limit}L. Bakemeier, A. Alvermann, and H. Fehske, Phys. Rev. A 85, 043821 (2012).
\bibitem{QFI}B. R. Frieden, Phys. Rev. A 41, 4265 (1990).
\bibitem{QCRB}C. W. Helstrom, Quantum Detection and Estimation Theory (Academic Press, New York, 1976); See also S. D. Personick, IEEE Trans. Inf. Theory 17, 240 (1971); H. P. Yuen and M. Lax, ibid. 19, 740 (1973); A. S. Holevo, Probabilistic and Statistical Aspects of Quantum Theory (North-Holland, Amsterdam, 1982).
\bibitem{qmtb}D. Xiao, M.-C. Chang, and Q. Niu, Rev. Mod. Phys. 82, 1959 (2010).
\bibitem{berry}X. Tan, D.-W. Zhang, Q. Liu, G. Xue, H.-F. Yu, Y.-Q. Zhu, H. Yan, S.-L. Zhu, and Y. Yu, Phys. Rev. Lett. 120, 130503 (2018).
\bibitem{chen1}M. D. Schroer, M. H. Kolodrubetz, W. F. Kindel, M. Sandberg, J. Gao, M. R. Vissers, D. P. Pappas, A. Polkovnikov, and K. W. Lehnert, Phys. Rev. Lett. 113, 050402 (2014).
\bibitem{chen2}P. Roushan, C. Neill, Y. Chen, M. Kolodrubetz, C. Quin-
tana, N. Leung, M. Fang, R. Barends, B. Campbell,
Z. Chen, B. Chiaro, A. Dunsworth, E. Jeffrey, J. Kelly,
A. Megrant, J. Mutus, P. J. J. O'Malley, D. Sank,
A. Vainsencher, J. Wenner, T. White, A. Polkovnikov,
A. N. Cleland, and J. M. Martinis, Nature 515, 241 (2014).
\bibitem{chen3}Z.-L. Zhang, M.-F. Chen, H.-Z. Wu, and Z.-B. Yang, Phys. Rev. D 95, 046010 (2017).
\bibitem{chen4}B. Wang, H. Zhang, S. Yang, F. Wu, Z. Yang, and S. Zheng, Adv. Quantum Technol. 6, 2300068 (2023).
\bibitem{qgt}P. Zanardi, P. Giorda, and M. Cozzini, Phys. Rev. Lett. 99, 100603 (2007).
\bibitem{HPT}T. Holstein and H. Primakoff, Phys. Rev. 58, 1098 (1940)
\end{thebibliography}
\end{document}